\documentclass[showpacs,preprintnumbers,amsmath,amssymb,nofootinbib]{revtex4}

\usepackage{epsf}
\usepackage{psfrag}
\usepackage{epsfig}
\usepackage{graphics}
\usepackage{amsmath,amssymb,graphicx}
\usepackage{slashed}
\usepackage[dvips]{color}

\begin{document}

\newcommand{\be}{\begin{equation}}
\newcommand{\ee}{\end{equation}}
\newcommand{\bq}{\begin{eqnarray}}
\newcommand{\eq}{\end{eqnarray}}
\newcommand{\dt}{\frac{d^3k}{(2 \pi)^3}}
\newcommand{\dtp}{\frac{d^3p}{(2 \pi)^3}}
\newcommand{\Dbruto}{\hbox{$D \!\!\!{\slash}$}}
\newcommand{\kbruto}{\hbox{$k \!\!\!{\slash}$}}
\newcommand{\pbruto}{\hbox{$p \!\!\!{\slash}$}}
\newcommand{\qbruto}{\hbox{$q \!\!\!{\slash}$}}
\newcommand{\lbruto}{\hbox{$l \!\!\!{\slash}$}}
\newcommand{\bbruto}{\hbox{$b \!\!\!{\slash}$}}
\newcommand{\parbruto}{\hbox{$\partial \!\!\!{\slash}$}}
\newcommand{\Abruto}{\hbox{$A \!\!\!{\slash}$}}
\newcommand{\dk}{\frac{d^4k}{(2 \pi)^4}}
\newcommand{\dl}{\frac{d^4l}{(2 \pi)^4}}
\newcommand{\pa}{\partial}
\def\pls{\partial\!\!\!/}
\def\bb{\bibitem}
\def\as{a\!\!\!/}
\def\As{A\!\!\!/}
\def\ks{k\!\!\!/}
\def\ls{l\!\!\!/}
\def\ps{p\!\!\!/}
\def\qs{q\!\!\!/}
\def\bs{b\!\!\!/}
\def\yb{\bar{\y}}
\def\Ds{D\!\!\!\!/}
\def\ds{\partial\!\!\!/}
\newcommand{\fr}{\frac}
\def\ov{\over}
\def\g{\gamma}
\def\n{\nu}
\def\m{\mu}
\def\n{\nu}
\def\bb{\bibitem}
\def\eff{\mathrm{eff}}
\def\CS{\mathrm{CS}}
\newcommand{\Slash}[1]{{#1}\!\!/}
\newcommand{\SLASH}[1]{{#1}\!\!\!/}
\newcommand{\RM}[1]{\mathrm{#1}}

\title{\textbf{Remarks on a Lorentz-breaking 4D chiral gauge theory}}

\date{\today}

\author{A. P. Ba\^eta Scarpelli$^{(a, b)}$} \email []{scarpelli.apbs@dpf.gov.br}
\author{M. Gomes$^{(b)}$} \email []{mgomes@if.usp.br}
\author{A. Yu. Petrov$^{(c)}$} \email[]{petrov@fisica.ufpb.br}
\author{A. J. da Silva$^{(b)}$} \email[]{ajsilva@if.usp.br}

\affiliation{(a) Setor T\'ecnico-Cient\'{\i}fico - Departamento de Pol\'{\i}cia Federal \\
Rua Hugo D'Antola, 95 - Lapa - S\~ao Paulo - Brazil}

\affiliation{(b) Instituto de F\'{\i}sica, Universidade de S\~ao Paulo,
Caixa Postal 66318, 05315-970, S\~ao Paulo, SP, Brazi}

\affiliation{(c) Departamento de F\'{\i}sica, Universidade Federal da Para\'{\i}ba\\
 Caixa Postal 5008, 58051-970, Jo\~ao Pessoa, Para\'{\i}ba, Brazil}

\begin{abstract}
We investigate a Lorentz-violating chiral model composed by two fermions, a complex scalar field and a gauge field. We show that by conveniently adjusting the parameters of the model, it is possible to generate an unambiguous Carroll-Field-Jackiw term and, at the same time, provide the  cancellation of the chiral anomaly. The renormalizability of the model is investigated and it is shown that the same counterterms needed in the symmetric phase also renormalize the model with broken symmetry.
\end{abstract}


\maketitle

\section{Introduction}

Symmetries play a fundamental role in nature. Physical concepts and conservation laws are deeply intertwined with symmetries of nature, as stated by the Noether theorem. However, symmetries sometimes impose such restrictions that, if they were exact, observed phenomena would be impossible. Thus, so important as the symmetries are their breaking mechanisms. Actually, it is desired that a Lagrangian density for a model be symmetric, although the world which it describes is not. This is the reason why one of the important subjects of research nowadays is the study of symmetry breaking mechanisms. In fact, one of the basic ingredients of the Standard Model (SM) is the so called Higgs mechanism, which is based on the spontaneous gauge symmetry breaking, which allows the generation of the masses of elementary particles, initially massless in the symmetric phase.

In some circumstances symmetries which occur at the classical level are broken in the quantum process. The cause of this are the so called anomalies and their presence may have deep consequences. For example, the quantum non-conservation of the chiral current, known as the chiral anomaly, opens the possibility for the theoretical explanation of the decay of a neutral pion in two photons. The chiral symmetry is considered as a global symmetry in some models. However, in the known chiral theories, which are part of the Standard Model of elementary particles, the local symmetry has a chiral component. In this way, the non conservation of the chiral current has, as a consequence, the breaking of the gauge symmetry of the theory. The quantum breaking of a local symmetry of the classical  model is harmful, causing the violation of unitarity and, consequently, spoiling the renormalizability of the model. Thus, for the consistency of chiral theories, it is necessary that the quantum anomalies are cancelled out. For this,  the model must encompass a set of fermions with chiral charges $Q_i$, such that they combine to cancel the anomaly. For example, for a set of left-handed fermions in an Abelian model, the condition is $\sum Q_i^3=0$. In general, the anomaly is odd in the chiral charge and if the fermions (in Abelian models) come in pairs of opposite axial charges, the condition for restoration of gauge symmetry is fulfilled \cite{Gross-Jackiw}.  It is instigating that the elementary particles which are part of the Standard Model all combine such that the chiral anomaly is cancelled out.

The Lorentz and CPT symmetries, which are basic for the construction of all modern phenomenological quantum field models, have been observed in all experimental tests \cite{data-exp}. However, even for these primordial symmetries, investigations are carried out to study the would be implications of small violations not yet experimentally discovered. In this context, the interest in Lorentz- and CPT-violating models  has increased since a Chern-Simons-like term in four dimensions was first considered by Carroll, Field and Jackiw \cite{jackiw}. This so called Carroll-Field-Jackiw (CFJ) term was included in a Standard Model Extension (SME), constructed to provide Lorentz and CPT violations, controlled by coefficients constrained by experiments \cite{kostelecky1}-\cite{coleman2}.

The investigation on the possibility that the CFJ term could be radiatively induced from the fermionic sector, whenever the axial term $b_{\mu }\bar{\psi} \gamma^{\mu }\gamma_{5}\psi$ is included, generated some controversy. The discussion were focused on some aspects: the first is related to the dependence of the induced term on the regularization scheme; the second is the possibility of imposing the vanishing of this term on physical grounds, like gauge invariance, causality and unitarity; and, finally, there is the question on whether or not the stringent limits on the magnitude of the coefficient of the CFJ term impose restrictions on the existence of the CPT- and Lorentz-breaking axial term in the fermionic sector. Many papers have been devoted to this subject (see, for example, the references \cite{CS1}-\cite{Gomes}).

An alternative model was, then, proposed, in which the CFJ term induced by quantum corrections has no ambiguities in its coefficient \cite{amb-free}. It is a chiral model, in which the background vector field $b_\mu$ and the gauge field couple with opposite chiralities to the fermion. This interesting model, however, needs some further development, since, for containing only one fermion, it is not capable of dealing with the gauge anomaly which is inherent to the chiral theories. Furthermore, the fermion mass part, which is necessary to the generation of the unambiguous Chern-Simons-like term, should be justified by a Higgs mechanism, since it explicitly violates the gauge symmetry of the model. In this paper, we study a more general Lorentz-violating chiral model composed of two fermions, a complex scalar field and a gauge field. The model respects a modified gauge symmetry. By means of the Higgs mechanism, the fermions and the gauge field acquire mass. We show that by adjusting certain coefficients, it is possible to generate an unambiguous CFJ term and, at the same time, provide the desired cancellation of the anomaly. The renormalizability of the model is investigated and it is shown that the same counterterms needed in the symmetric phase also renormalize the model with broken symmetry. Although this is a well-known result for Lorentz invariant theories, it has not been deeply investigated in the case of Lorentz-breaking models. As we will see, a Lorentz-violating part must be included in the covariant derivative of the scalar field in order to preserve this feature at one-loop order.

This work is organized as follows. In section II, the model is presented. Since the local gauge symmetry is spontaneously broken, the complex scalar field is decomposed in two real fields and the final Lagrangian density is written. We obtain the condition to the induction of an unambiguous CFJ term in section III. In section IV, the cancellation of the anomaly is discussed. The renormalization of the model is discussed in section V. In section VI, we summarize our results and present some concluding comments. In appendix A, the details of the one-loop renormalization are shown and the results of individual graphs of appedndix A are presented in appendix B.

\section{The Lorentz-violating chiral model}

We begin with the following Lorentz-violating model,
\bq
&&\mathcal L= -\frac 14 F_{\mu \nu}F^{\mu \nu} + (D_\mu \phi)^*(D^\mu \phi) + \mu^2 \phi^* \phi - \frac{\lambda}{4} (\phi^* \phi)^2 +\nonumber \\
&& +\bar \psi\left[i\pls -\bs (C P_L+D P_R) -e \As (M P_L +N P_R) \right]\psi  +\nonumber \\
&&+\bar \chi\left[i\pls -\bs (C' P_L+D' P_R) -e \As (N P_L +M P_R) \right]\chi +\nonumber \\
&& -g \left( \bar \psi_L \psi_R \phi + \bar \psi_R \psi_L \phi^*\right)
-g' \left( \bar \chi_R \chi_L \phi + \bar \chi_L \chi_R \phi^*\right),
\label{model1}
\eq
where $\psi$ and $\chi$ are Dirac fermions and
\be
\psi_{R,L}=P_{R,L} \psi,
\ee
being
\be
P_{R,L}=\frac{1\pm \gamma_5}{2}
\ee
the chiral projectors. Besides, $C$, $D$, $C'$, $D'$, $M$ and $N$ are real constants and $b_\mu$ is a constant four-vector which breaks the Lorentz symmetry. The covariant derivative acting on the complex scalar field is given by
\be
D_\mu \phi = \left(\pa_\mu + i \kappa b_\mu + i e' A_\mu \right)\phi,
\ee
in which $\kappa$ is a dimensionless constant. The introduction of a Lorentz-violating sector in the covariant derivative is necessary for the closure of the model when radiative corrections are considered. This will be evident in the study of the renormalizability of the model. The parameter $\kappa$ will allow us to follow what happens when we turn off this part in the covariant derivative. The Lagrangian density of eq. (\ref{model1}) is invariant under the local transformations
\bq
\label{GT}
&&\psi \to e^{-i e\alpha(x)  (M P_L+N P_R)}\psi, \nonumber \\
&& \bar \psi \to \bar \psi e^{i e\alpha(x)( N P_L+ M P_R)}, \nonumber \\
&&\chi \to e^{-ie \alpha(x) (N P_L+M P_R)}\chi, \nonumber \\
&& \bar \chi \to \bar \chi e^{i e\alpha(x)( M P_L+ N P_R)}, \nonumber \\
&& \phi \to  e^{-i e' \alpha(x)} \phi,  \nonumber \\
&& \phi^* \to  e^{i e' \alpha(x)} \phi^*  \,\,\,\, \mbox{and} \nonumber \\
&&A_\mu \to A_\mu- \pa_\mu \alpha (x),
\eq
with $e'=(M-N)e$.

This model, as we will see in more details in section IV, avoids the problem of the gauge anomaly, since the two fermionic fields, $\psi$ and $\chi$, posses opposite chiral charges \cite{Gross-Jackiw}. The Lagrangian density (\ref{model1}), if the covariant derivative is written explicitly, takes the form,
\bq
&&\mathcal L= -\frac 14 F_{\mu \nu}F^{\mu \nu} + (\pa_\mu \phi^*)(\pa^\mu \phi)+
\mu'^2 \phi^* \phi - \frac{\lambda}{4} (\phi^* \phi)^2 + e'^2A_\mu A^\mu\phi^* \phi +\nonumber \\
&& +ie' A^\mu\left(\phi \pa_\mu \phi^*-\phi^* \pa_\mu \phi\right) + 2e' \kappa b_\mu A^\mu \phi^* \phi
+ i \kappa b^\mu \left(\phi \pa_\mu \phi^*-\phi^* \pa_\mu \phi\right)+\nonumber \\
&&+\bar \psi\left[i\pls -\bs (C P_L+D P_R) -e \As (M P_L +N P_R) \right]\psi + \nonumber\\
&&+\bar \chi\left[i\pls -\bs (C' P_L+D' P_R) -e \As (N P_L +M P_R) \right]\chi +\nonumber \\
&&-g \left( \bar \psi_L \psi_R \phi + \bar \psi_R \psi_L \phi^*\right)
-g' \left( \bar \chi_R \chi_L \phi + \bar \chi_L \chi_R \phi^*\right).
\label{model2}
\eq

Rewriting the Lagrangian (\ref{model2}) in terms of real scalar fields $\rho$ and $\varphi$, such as $\phi=2^{-\frac{1}{2}}\left(\rho+v+i\varphi\right)$, being $v$ a constant, we obtain
\be
\mathcal L=\mathcal L_\phi+\mathcal L_A + \mathcal L_\psi + \mathcal L_\chi,
\label{total}
\ee
with
\bq
&&\mathcal L_\phi= \frac 12 (\pa_\mu \rho)^2 + \frac 12 (\pa_\mu \varphi)^2 - \frac12 m_\rho^2 \rho^2
+ \frac{e'^2}{2}A_\mu A^\mu\left(\rho^2+\varphi^2+2v\rho+ v^2\right) +\nonumber \\
&&+ e'\kappa b_\mu A^\mu \left(\rho^2+\varphi^2+2v\rho+ v^2\right) + e' A^\mu \left(\rho \pa_\mu \varphi-\varphi \pa_\mu \rho\right) + e v A^\mu \pa_\mu \varphi
+ \kappa b^\mu \left(\rho \pa_\mu \varphi-\varphi \pa_\mu \rho\right) + \nonumber \\
&& - \frac{\lambda}{16}\left(\rho^2+\varphi^2\right)^2 -\frac{\lambda}{4}v \rho \left(\rho^2+\varphi^2\right)
-\frac 12 \delta m^2\left(\rho^2+\varphi^2+2v \rho\right),
\label{Lphi}
\eq
\bq
\mathcal L_A= -\frac 14 F_{\mu \nu}F^{\mu \nu}- \frac {1}{2\xi}\left(\pa_\mu A^\mu\right)^2,
\label{La}
\eq
\bq
\mathcal L_\psi=\bar \psi\left[i\pls -m_\psi -\bs (C P_L+D P_R) -e \As (M P_L +N P_R)
- \frac{g}{\sqrt{2}}\rho -i \frac{g}{\sqrt{2}}\gamma_5 \varphi\right]\psi
\label{Lpsi}
\eq
and
\bq
\mathcal L_\chi=\bar \chi\left[i\pls -m_\chi -\bs (C' P_L+D' P_R) -e \As (N P_L +M P_R)
- \frac{g'}{\sqrt{2}}\rho +i \frac{g'}{\sqrt{2}}\gamma_5 \varphi\right]\chi,
\label{Lpsi}
\eq
where $\delta m^{2}=-\mu'^2+\frac{\lambda}{4} v^{2}$, $m_{\rho}^{2}=\frac{\lambda v^{2}}{2}$, $m_\psi=\frac{gv}{\sqrt{2}}$, $m_\chi=\frac{g'v}{\sqrt{2}}$, $m_{A}^{2}=v^{2}e^{2}$ and we have added a gauge fixing term, $\mathcal L_{GF}=-\left(2\xi\right)^{-1}\left(\partial_{\mu}A^{\mu}\right)^{2}$.

For $\lambda$ and $\mu'^2=\mu^2 + \kappa^2 b^2$ real and positive\footnote{Actually, $b^2$ can be negative. However, for small Lorentz violations, it is reasonable to require that $\mu^2>|\kappa^2 b^2|$.}, the complex scalar field $\phi$ develops a non null vacuum expectation value $\langle\phi\rangle_{0}=\frac{v}{\sqrt{2}}$, with $\mu'^2=\frac{\lambda}{4}v^2$ and, so, the local $U(1)$ symmetry is spontaneously broken. The model presented above, which is power counting renormalizable, will be carefully investigated in the next sections. We begin by establishing conditions to have an ambiguity-free induced CFJ term.

\section{The ambiguity-free CFJ term}

In this section, we calculate the one-loop vacuum polarization tensor at first order in $b_\mu$ to verify under which conditions, the induced CFJ term is ambiguity free. It is natural to conclude that the unique possibility of radiatively generating this term is by means of fermion loops, since the desired term encompasses an antisymmetric  Levi-Civita tensor. This is because only the fermionic sector contains an axial part (proportional to $\gamma_5$). There are two approaches which could be carried out to extract the first order in $b_\mu$. The first one consists of taking the $\psi$ and $\chi$ fermion loops and then expanding the propagators as
\be
S_\psi(k)=\frac{i}{\ks-m_\psi -\bs (C P_L+ D P_R)}= \sum_{n=0}^{\infty}\frac{i}{\ks-m_\psi}\left[-i\bs (C P_L+ D P_R) \frac{i}{\ks-m_\psi}\right]^n \equiv \sum_{n=0}^{\infty} S_n(k),
\ee
to consider only the first order in $b_\mu$,
\be
S_\psi(k)\approx S_0(k)+S_1(k).
\ee

\begin{figure}[h!]
\unitlength1cm
\centerline{\hbox{
     \epsffile{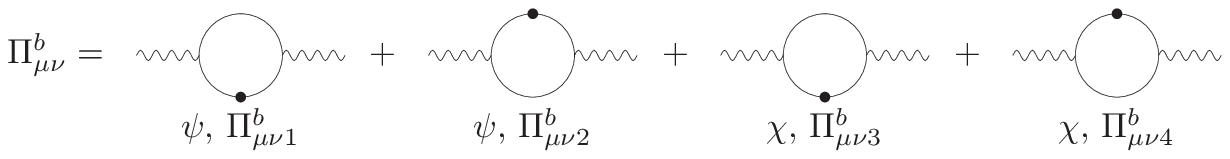}
     }
  }
\caption[]{Contributions to the CFJ term. In the two first graphs, we have a loop of the $\psi$ fermion, whereas the other two graphs have a loop of the fermion $\chi$.}
\label{CFJ-contributions}
\end{figure}

The second method consists of considering the $b_\mu$-dependent terms as interactions and taking only the contributions with one insertion of these interactions. Following this second approach, the graphs which contribute to the CFJ term are depicted in Fig. \ref{CFJ-contributions}, in which the dots indicate insertions of $b_\mu$.

We now calculate in details the contribution of the first graph,
\be
\Pi^b_{\mu \nu 1}=-e^2 \int \frac{d^4k}{(2\pi)^4}\frac{N_{\mu \nu}}{(k^2-m_\psi^2)^2[(k+p)^2-m_\psi^2]},
\ee
with
\be
N_{\mu \nu}=\mbox{tr}\left\{\gamma_\mu (M P_L+N P_R)(\ks+m_\psi)\bs (C P_L+ D P_R)(\ks+m_\psi)\gamma_\nu (M P_L+N P_R)(\ks+\ps+m_\psi)\right\}.
\ee
Since we are interested in the Chern-Simons-like term, we will only consider here the terms which can yield the Levi-Civita symbol. After carrying out the traces of products of Dirac matrices in four dimensions, we obtain
\bq
N_{\mu \nu}&=& 2i\varepsilon_{\mu \nu \alpha \beta}\left\{-b^\alpha k^\beta\left[(M^2C-N^2D)k^2+(N^2C-M^2D+2MN(D-C))m_\psi^2 \right] +\right.\nonumber\\
  &-& \left.b^\alpha p^\beta \left[ (M^2C-N^2D)k^2+(N^2C-M^2D)m_\psi^2 \right]+2(b \cdot k)k^\alpha p^\beta(M^2C-N^2D)\right\}+ \nonumber \\
 &+& N'_{\mu \nu},
\eq
where $N'_{\mu \nu}$ represents the terms which do not involve the Levi-Civita tensor. We split the result in two parts,
\be
\Pi^b_{\mu \nu 1}=\Pi^5_{\mu \nu 1}+\Pi'^b_{\mu \nu 1},
\ee
being $\Pi^5_{\mu \nu 1}$ the contribution to the CFJ term. So, we have
\bq
\Pi^5_{\mu \nu 1}&=&-2ie^2 \varepsilon_{\mu \nu \alpha \beta}\left\{ -b^\alpha\left[(M^2C -N^2D)J'^\beta + (N^2C-M^2D+2MN(D-C))m_\psi^2J^\beta+ \right. \right. \nonumber \\
&-& \left. \left. p^\beta\left((M^2C -N^2D)J' + (N^2C-M ^2D)m_\psi^2 J\right)\right]+ 2 b_\rho p^\beta (M^2C -N^2D) J^{\alpha \rho} \right\},
\eq
in which we have defined the integrals
\bq
J, J^\beta, J^{\alpha \beta}&=& \int \frac{d^4k}{(2\pi)^4}\frac{1, k^\beta, k^\alpha k^\beta}{(k^2-m_\psi^2)^2[(k+p)^2-m_\psi^2]}, \\
J', J'^\beta&=& \int \frac{d^4k}{(2\pi)^4}\frac{k^2, k^2 k^\beta}{(k^2-m_\psi^2)^2[(k+p)^2-m_\psi^2]}.
\eq

In the above result, the integrals $J^{\alpha \beta}$, $J'$ and $J'^\beta$ are divergent. We would like to show that the total result, considering all the graphs of Fig \ref{CFJ-contributions}, can be unambiguous for some choice of the coefficients $M$, $N$, $C$, $D$, $C'$ and $D'$. To make this evident, at least in this section we will use an approach which does not resort explicitly to any particular regularization procedure. For dealing with the divergent integrals, we use recursively the identity
\be
\frac{1}{(p+k)^2-m^2}=\frac{1}{k^2-m^2}-\frac{p^2+2p\cdot k}{(k^2-m^2)\left[(p+k)^2-m^2\right]},
\ee
in order to extract the external momentum, $p$, from the divergent integrals. We then employ the implicit regularization approach based on extracting the surface terms (for a description of this method, see f.e. \cite{Adriano}). Following this procedure, we obtain, for the first graph,
\be
\Pi^5_{\mu \nu 1}=-2i e^2 \varepsilon_{\mu \nu \alpha \beta} b^\alpha p^\beta\left\{ \frac{i}{48 \pi^2}(M^2+N^2+MN)(C-D)-(M^2C-N^2D)\alpha\right\},
\ee
where the surface term
\be
\alpha g_{\mu \nu}\equiv g_{\mu \nu}\int \frac{d^4k}{(2\pi)^4}\frac{1}{(k^2-m_\psi^2)^2}- 4\int \frac{d^4k}{(2\pi)^4}\frac{k_\mu k_\nu}{(k^2-m_\psi^2)^3}
\ee
is responsible for the regularization dependence of this first contribution and the limit $p^2 \to 0$ was taken. It should be noticed that the mass $m_\psi^2$ in the definition of $\alpha$ can be replaced by an arbitrary mass scale, since the finite final result will not depend on $m_\psi^2$ in any regularization prescription.

Adopting the same procedures for the other three graphs, one obtains
\be
\Pi^5_{\mu \nu 2}=-i e^2 \varepsilon_{\mu \nu \alpha \beta} b^\alpha p^\beta\left\{ \frac{i}{24 \pi^2}(M^2+N^2+MN)(C-D)\right\},
\ee
\be
\Pi^5_{\mu \nu 3}=-2i e^2 \varepsilon_{\mu \nu \alpha \beta} b^\alpha p^\beta\left\{ \frac{i}{48 \pi^2}(N^2+M^2+MN)(C'-D')-(N^2C'-M^2D')\alpha\right\}
\ee
and
\be
\Pi^5_{\mu \nu 4}=-i e^2 \varepsilon_{\mu \nu \alpha \beta} b^\alpha p^\beta\left\{ \frac{i}{24 \pi^2}(N^2+M^2+MN)(C'-D')\right\},
\ee
so that
\be
\Pi^5_{\mu \nu}=e^2 \varepsilon_{\mu \nu \alpha \beta}b^\alpha p^\beta \left\{2i\alpha \left[M^2(C-D')-N^2(D-C')\right] + \frac {1}{12 \pi^2}(M^2+N^2+MN)(C+C'-D-D')\right\}.
\label{CFJ-result}
\ee

Examining the final result for the Chern-Simons-like term above, we can find that it represents itself an undetermined number of combinations of the coefficients $M$, $N$, $C$, $D$, $C'$ and $D'$, whose relation can be fixed to obtain an $\alpha$-independent and, hence, finite and unambiguous result. If we take, for example, $C-D'=4(D-C')$ and $N=2M$, we have
\be
\Pi^5_{\mu \nu}=\frac {7 e^2}{4 \pi^2} \varepsilon_{\mu \nu \alpha \beta}b^\alpha p^\beta M^2(D-C').
\ee

Another interesting peculiarity of the result in eq. (\ref{CFJ-result}) is the possibility of the vanishing of the unambiguous part, if the condition $C+C'=D+D'$ is satisfied. It is worth to understand what happens in this situation. A particular solution is $C=D$ and $C'=D'$. In this case, the axial part of the coupling of the background vector $b_\mu$ with the fermions is zero. The remaining coupling is of the type $-\bar \psi \bs \psi$, which can be absorbed in a redefinition of the fields. For this particular case, it is expected that, even at higher loop orders, the unambiguous part of the CFJ term is not induced. Nevertheless, there is an infinite number of solutions which maintain the chiral part of the coupling of $b_\mu$ with the fermions. We conjecture that in this situation the one-loop cancellation of the unambiguous sector is only casual and probably does not hold at higher loop orders.

In the next section we will discuss the cancellation of the anomaly.

\section{Anomaly cancellation}

An important issue related with chiral models is the problem of the axial anomaly. Since the local symmetry of such models includes a chiral component, the anomaly can have undesirable consequences, like violation of unitarity and destruction of the renormalizability of the theory. Let us discuss how our model is constructed in order to provide the desired cancellation of the anomaly. For this discussion, we consider the model in the symmetric phase and without the presence of the complex scalar field (for the case of broken symmetry phase, the scalar fields, $\rho$ and $\varphi$, should be taken into account). Let us also initially consider only one fermion and write down the field equations for $\psi$ and $\bar \psi$,
\be
\left[ i \ds -\bs (C P_L+ D P_R) - e\As (M P_L+ N P_R)\right]\psi=0
\ee
and
\be
\bar \psi\left[i\overleftarrow \ds +\bs (C P_L+ D P_R) + e\As (M P_L+ N P_R)\right]=0.
\ee
The most simple current to be constructed is the vectorial one, given by $j_\mu=e\bar \psi \gamma_\mu \psi$, such that
\be
\pa^\mu j_\mu=e \bar \psi \overleftarrow \ds \psi + e \bar \psi \ds \psi,
\ee
in which the field equations are to be used. It is straightforward to obtain $\pa^\mu j_\mu=0$. Let us consider now the chiral current, $j_\mu^5=e \bar \psi \gamma_\mu \gamma_5 \psi$. We get
\be
\pa^\mu j_\mu^5=e \left( \bar \psi \overleftarrow \ds \gamma_5 \psi - \bar \psi \gamma_5 \ds \psi \right)=0,
\ee
where again we made use of the field equations. We can further combine the two currents in order to get
\be
j_\mu^T=e \bar \psi \gamma_\mu (M P_L+ N P_R)\psi,
\ee
so that
\be
\pa^\mu j_\mu^T=0.
\ee

\begin{figure}[h!]
\unitlength1cm
\centerline{\hbox{
     \epsffile{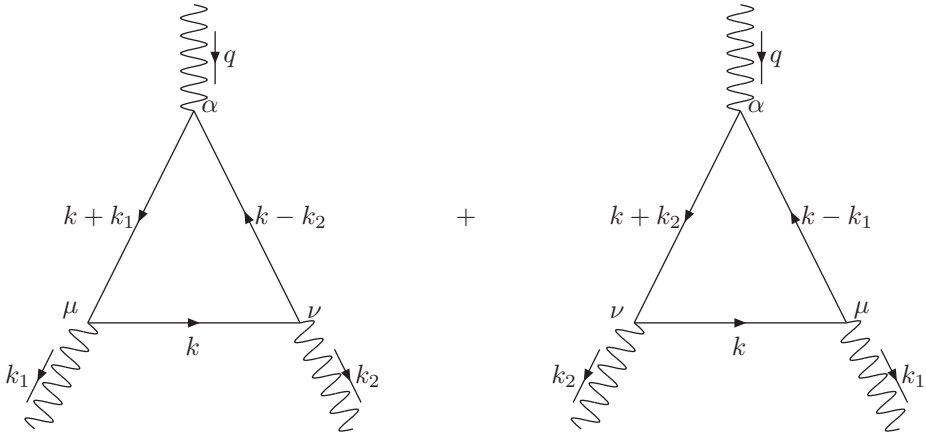}
     }
  }
\caption[]{Contributions to the vertex $V_{\mu \nu \alpha}$}
\label{anomaly}
\end{figure}

The conservation of the current $J_\mu^T$ is the one that is required by the local symmetry as exposed in the transformations (\ref{GT}). If one considers a triangle graph with three external photons, as depicted in Fig. \ref{anomaly}, based on the classical symmetry, one can expect that
\bq
&& q^\alpha V_{\mu \nu \alpha}=0, \\
&& k_1^\mu V_{\mu \nu \alpha}=0, \\
&& k_2^\nu V_{\mu \nu \alpha}=0.
\eq

Before continuing, it is important to remember that the anomaly cancellation of the Lorentz invariant version of the present model has been shown in \cite{Gross-Jackiw}. In fact, there is no difference here. The superficial degree of divergence of the Feynman graphs in our model is given by
\be
\mathcal{D}= 4-N_b-\frac 32 N_f - \sum_i c_i V^{SR}_i,
\ee
where $N_b$ is the number of external boson lines, $N_f$ is the number of external fermion lines and $V^{SR}_i$ are the number of super-renormalizable vertices with the coefficients $c_i$, which, depending on the case, is $1$ or $2$. The insertion of a $b_\mu$ in one of the fermion lines of Fig. \ref{anomaly} will turn the integral logarithmically divergent and, so, the Lorentz-breaking vector will not contribute to the anomaly. The discussion which follows is then restricted to the zeroth order in $b_\mu$ and is just illustrative, since it matches perfectly that one of \cite{Gross-Jackiw}.

We have for the graphs of Fig. \ref{anomaly},
\be
V_{\mu \nu \alpha}=e^3  \int \frac{d^4k}{(2\pi)^4} \frac{N_{\mu \nu \alpha}}{k^2(k+k_1)^2(k-k_2)^2} +\,\,\,\mbox{crossed graph},
\ee
with
\be
N_{\mu \nu \alpha}=\mbox{tr}\left\{\gamma_\nu(M P_L+N P_R)\ks \gamma_\mu (M P_L+N P_R)(\ks+\ks_1)\gamma_\alpha (M P_L+N P_R)(\ks -\ks_2)\right\}.
\ee
The chiral projectors can be moved through the gamma matrices, using that $P_L \gamma_\mu=\gamma_\mu P_R$ and $P_R \gamma_\mu=\gamma_\mu P_L$. So, we obtain
\bq
N_{\mu \nu \alpha}&=&\mbox{tr}\left\{\gamma_\nu \ks \gamma_\mu (\ks+\ks_1)\gamma_\alpha (M^3P_L+N^3P_R)(\ks-\ks_2)\right\} \\ \nonumber
&=& \frac{M^3+N^3}{2}\mbox{tr}\left\{\gamma_\nu \ks \gamma_\mu (\ks+\ks_1)\gamma_\alpha (\ks-\ks_2)\right\}
+ \frac{N^3-M^3}{2}\mbox{tr}\left\{\gamma_\nu \ks \gamma_\mu (\ks+\ks_1)\gamma_\alpha \gamma_5(\ks-\ks_2)\right\},
\eq
which will give
\be
V_{\mu \nu \alpha}=e^3 \left(\frac{M^3+N^3}{2}V_{\mu \nu \alpha}^{(1)}+ \frac{N^3-M^3}{2} V_{\mu \nu \alpha}^{(2)} \right)\,\, + \,\, \mbox{crossed diagram},
\ee
where $V_{\mu \nu \alpha}^{(1)}$ is the triangle with three vectorial vertices and $V_{\mu \nu \alpha}^{(2)}$ is the triangle with one axial and two vectorial vertices. The first part will give a zero result when contracted with the external momenta, whereas the second part gives the axial anomaly. Before addressing the other fermionic field, we rewrite the coupling of $\psi$ to the gauge field as
\be
-e\bar \psi\As (M P_L +N P_R)\psi= -e\bar \psi \As \left(\frac{M+N}{2}+\frac{N-M}{2}\gamma_5\right) \psi
\ee
and identify $f= e\frac{M+N}{2}$ as the vectorial charge and $g=e\frac{N-M}{2}$ as the axial charge. In this way, we can see that the coefficient of the anomaly is given by
\be
\frac{N^3-M^3}{2}=g(g^2+3f^2),
\ee
which is odd in the chiral charge and coincides with the result of \cite{Gross-Jackiw}.

We now consider the other fermion, $\chi$, which couples to the gauge field with opposite chiral charge. Since the coefficients $M$ and $N$ are exchanged in its Lagrangian density, it is easy to see that the anomaly is cancelled out when the two fermions are considered together.

\section{Renormalization}

Lorentz-invariant theories with spontaneous symmetry breaking are known to be renormalized, in the broken phase, by the same counterterms of the symmetric phase. However, there is not a general proof which includes the case where the Lorentz and CPT symmetries are broken. For the present model, this is not an obvious issue. As presented in section II, we included a Lorentz-violating part in the covariant derivative of the complex scalar field. In a first view, one could consider this unnecessary. However, when the one-loop corrections are computed, it is found that new terms arise which are not present in the Lagrangian without the mentioned contribution from the covariant derivative. This will be evident in the appendix A, in which we carry out the one-loop renormalization of the present model.

\section{Concluding comments}

We studied a Lorentz-breaking chiral model which have the interesting particularity of allowing the quantum induction of a finite ambiguity-free Carroll-Field-Jackiw term and which is a generalization of the model presented in the reference \cite{amb-free}. The present model, which includes two fermionic fields with opposite chiral charges and a Higgs sector, provides the cancellation of the gauge anomaly, which would be harmful for the unitarity and the renormalizability of the model. Furthermore, the Higgs field provides the desired mechanism allowing to generate the fermionic masses which are necessary for the induction of the ambiguity-free Chern-Simons-like term.

The one-loop renormalization was also studied. In this respect, there are some subtleties in the model, which is power counting renormalizable. For consistency of the renormalization, a Lorentz-violating part was included in the covariant derivative of the complex scalar field. This covariant derivative provides exactly the terms that would be lacking for the one-loop renormalization of the model. All superficially divergent one-loop graphs with dependence in the Lorentz-breaking background vector were computed and we showed that the same counterterms that renormalize the theory in the symmetric phase are needed after the Higgs mechanism takes place. This is an example of a consistent Lorentz-violating chiral model.

It was shown in \cite{ouranom} that new contributions to the chiral anomalies depending on the Lorentz violating parameters cannot arise in absence of the term $\bar \psi \bs \gamma_5 \psi$. So, it is natural to expect that
the mechanism for cancellation of the chiral anomaly presented in this paper works also in other possible Lorentz-breaking extensions of QED, including the nonminimal ones \cite{ouranom}.
We are planning to discuss this issue in a forthcoming paper.

\vspace{2mm}
{\bf Acknowledgements.} This work was partially supported by Conselho Nacional de Desenvolvimento Cient\'{\i}fico e Tecnol\'{o}gico (CNPq). The work by A. Yu. P. has been partially supported by the CNPq project No. 303438/2012-6. A. P. B. S. thanks Pedro R. S. Gomes for useful discussions and the Physics Institute of Universidade de S\~ao Paulo for the hospitality.

\section{Appendix A - One-loop renormalization}

The Lorentz-invariant version of the model (\ref{total}) is renormalizable at all loop orders, as shown in \cite{Gross-Jackiw}. Thus, it remains to investigate the terms which depend on the background vector $b_\mu$. In what concerns the $b_\mu$ dependent terms, we adopt  a procedure  similar to the one used in \cite{Cleber}. The redefinition of the complex scalar field is such that the vacuum expectation value $\langle\rho\rangle_{0}$ of the field $\rho$ should vanish at the classical level, that is, $\delta m^{2}=0$. This gives the $\rho$ field a mass $m_{\rho}$. It is well known that the field $\varphi$ is the Goldstone boson. Actually, we can fix $\delta m^{2}$ as a counterterm to each order of perturbation theory using the normalization condition
\begin{equation}
\langle\rho\rangle_{0}=0 \label{eq:normcond1}
\end{equation}
at some renormalization scale.

In traditional theories, the renormalization of models in which the symmetry has been spontaneously broken is carried out with the same counterterms used in the original model in the symmetric phase. We follow the same procedure here. There are groups of terms in the Lagrangian density in which the gauge symmetry is broken which are generated from the same term of the Lagrangian of the model in the symmetric phase. So, they must renormalize together, with just one counterterm. We will treat each one of these groups separately.

\subsection{The first group}
We begin with the group
\be
e'\kappa b_\mu A^\mu \left(\rho^2+\varphi^2+2v\rho+ v^2\right),
\ee
which was generated from the term $2e' \kappa b_\mu A^\mu \phi^* \phi$. We define the counterterm
\be
\mathcal L_{CT-1}=2\delta_1 \kappa b_\mu A^\mu \phi^* \phi,
\ee
so that the following relation,
\be
v F_\mu^{A \rho \, \Lambda}=v^2\Gamma_\mu^{A\rho \rho \, \Lambda}= v^2 \tilde \Gamma_\mu^{A \varphi \varphi \, \Lambda}= T_\mu^{A \, \Lambda},
\label{fg}
\ee
between the divergent parts, indicated by the index $\Lambda$, of the corrections to the $A\rho$ line, to the $A \rho \rho$ and $A \varphi \varphi$ vertices and to the $A$ tadpole, respectively, must be respected. We remember that, although $v$ is constant, it should be taken as a background field. It should be noticed that, although some of the terms in the first group have different coefficients, when the symmetry factors of the counterterms are taken into account, we obtain the condition (\ref{fg}). This observation is important also for the other groups of terms.

\begin{figure}[h!]
\unitlength1cm
\centerline{\hbox{
     \epsffile{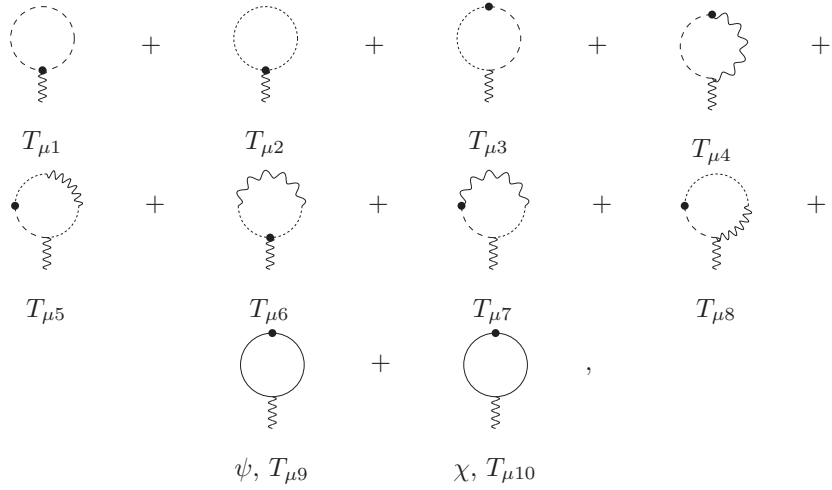}
     }
  }
\caption[]{One-loop contributions to the one-point function in $A_\mu$ at first order in $b_\mu$.}
\label{Tad-A}
\end{figure}

For the one-loop $A_\mu$ tadpole, we have the $b_\mu$-dependent divergent contributions given by Fig. \ref{Tad-A}, where the continuous lines represent fermions, the dashed lines represent the Higgs field $\rho$, the dotted lines represent the Goldstone field $\varphi$ and the wavy lines stand for the photon. The vertices represented by a big dot indicate where the Lorentz violating vector $b_\mu$ is inserted. These graphs are all of first order in $b_\mu$. There are also superficially divergent graphs of higher order in the Lorentz-violating parameter. However, they all either vanish or cancel out.
The individual results, considering only the divergent contributions, where the Feynman gauge ($\xi=1$) has been adopted, are displayed in appendix B.
The integrals can be solved, for example, by Dimensional Reduction, in which only the integrals are extended to a dimension $d$. We obtain the following result for the divergent part of the sum of the graphs, with $\epsilon=4-d$,
\be
T_\mu^\Lambda= -\frac{i}{8 \pi^2}v^2 b_\mu \left\{2\kappa e'^3+eg^2(M-N)(D-C)+eg'^2(N-M)(D'-C')\right\}\frac{1}{\epsilon}.
\ee

\begin{figure}[h!]
\unitlength1cm
\centerline{\hbox{
     \epsffile{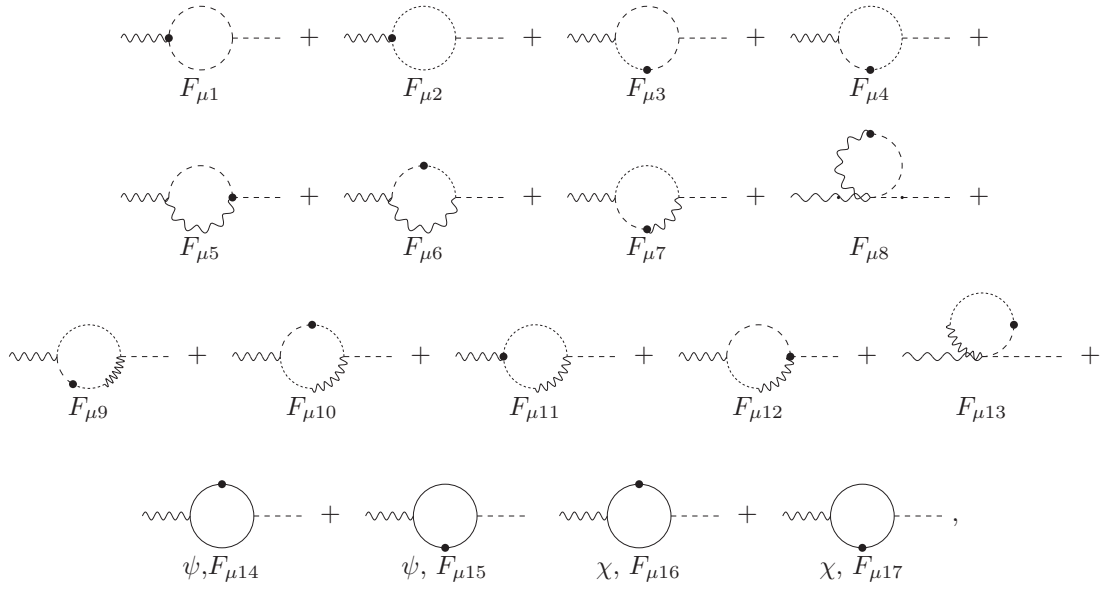}
     }
  }
\caption[]{One-loop contributions to the two-point function $\rho A$ at first order in $b_\mu$.}
\label{Rho-A}
\end{figure}

Considering the corrections to the $A\rho$ line, we have the superficially divergent contributions given by Fig. \ref{Rho-A}, which are also linear in $b_\mu$.
All the graphs together give us the divergent part
\be
F_\mu^\Lambda=-\frac{i}{8 \pi^2}v b_\mu \left[2\kappa e'^3+eg^2(M-N)(D-C)+eg'^2(N-M)(D'-C')\right]\frac{1}{\epsilon}.
\ee

\begin{figure}[h!]
\unitlength1cm
\centerline{\hbox{
     \epsffile{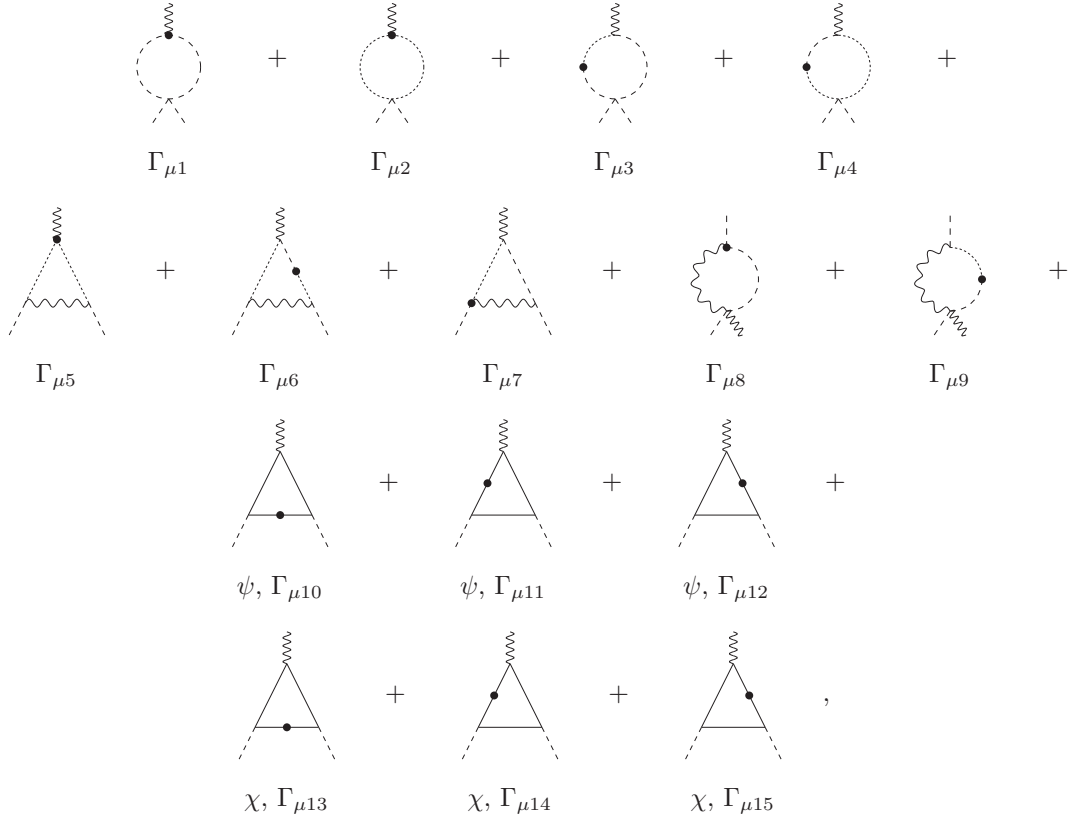}
     }
  }
\caption[]{One-loop contributions to the vertex $A \rho \rho$ at first order in $b_\mu$.}
\label{Rho-Rho-A}
\end{figure}

For the correction to the vertex $A\rho \rho$, the divergent graphs which depend on the background vector are shown in Fig \ref{Rho-Rho-A}.
Collecting all the terms, we get
\be
\Gamma_\mu^\Lambda=-\frac{i}{8 \pi^2}b_\mu \left[2\kappa e'^3+eg^2(M-N)(D-C)+eg'^2(N-M)(D'-C')\right]\frac{1}{\epsilon}.
\ee
Finally, for the $A \varphi \varphi$ vertex we have the same graphs of the $A \rho \rho$ vertex, with the replacement of the $\rho$ lines with the $\varphi$ lines, giving the same result:
\be
\tilde \Gamma_\mu^\Lambda=-\frac{i}{8 \pi^2}b_\mu \left[2 \kappa e'^3+eg^2(M-N)(D-C)+eg'^2(N-M)(D'-C')\right] \frac{1}{\epsilon}.
\ee

Thus, we checked that the divergent parts of the one-loop Green functions of this first group are equal, as written in eq. (\ref{fg}). Consequently, we verified that the counterterm that renormalizes the original term in the symmetric model, which gives origin to this first group, also renormalizes the four terms of the group in the model with broken symmetry.

\subsection{The second group}

We now turn our attention to the second group,
\be
-\frac 12 \delta m^2\left(\rho^2+\varphi^2+2v \rho\right),
\label{mass}
\ee
which is related to the preservation of the zero vacuum expectation value of the Higgs field. This group should renormalize together, being $\delta m^2$ the common coefficient of the counterterms expressed by eq. (\ref{mass}). Thus, it is expected that the divergent parts of the Higgs self-energy, the Goldstone self-energy and the Higgs tadpole respect the relation
\be
v \Sigma_{\rho \rho}^\Lambda=v \tilde \Sigma_{\varphi \varphi}^\Lambda= T_\rho^\Lambda.
\label{mass2}
\ee
Again, we observe that the above relation holds because of the symmetry factors in the counterterms.
\begin{figure}[h!]
\unitlength1cm
\centerline{\hbox{
     \epsffile{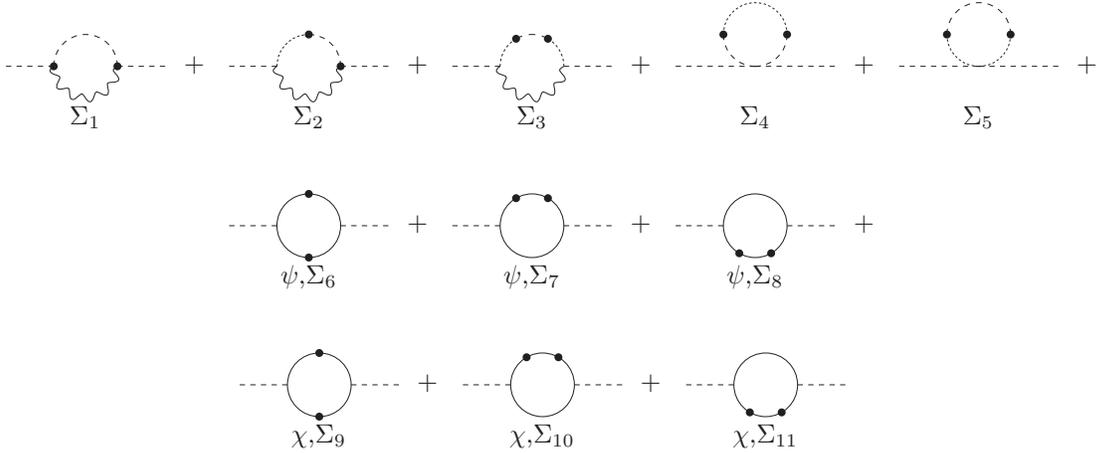}
     }
  }
\caption[]{Divergent one-loop contributions to the $\rho$ self-energy with dependence on $b_\mu$.}
\label{Rho-Rho}
\end{figure}
We begin with the $\rho$ self-energy. For this two-point function, the superficial degree of divergence is given by $\mathcal{D}=2-\sum c_i V_i^{SR}$. Since the vertices which contain $b_\mu$ are all super-renormalizable, we have divergent contributions only until second order in $b_\mu$. In first order, the graphs combine to give a zero divergent part. In second order, the contributions are given by the graphs displayed in Fig. \ref{Rho-Rho}. Collecting all the terms, we obtain
\be
\Sigma^\Lambda=\frac{i}{16 \pi^2}b^2\left[-g^2(C-D)^2-g'^2(C'-D')^2+\kappa^2 \lambda-3 \kappa^2 e'^2\right] \frac{1}{\epsilon}.
\ee

Next, we consider the Goldstone field self-energy. The graphs are similar to the ones of Fig. \ref{Rho-Rho}, the unique difference being the exchange of the $\rho$ lines with the $\varphi$ lines. The terms together give the same result,
\be
\tilde \Sigma^\Lambda=\frac{i}{16 \pi^2}b^2\left[-g^2(C-D)^2-g'^2(C'-D')^2+ \kappa^2 \lambda-3 \kappa^2 e'^2\right] \frac{1}{\epsilon}.
\ee

\begin{figure}[h!]
\unitlength1cm
\centerline{\hbox{
     \epsffile{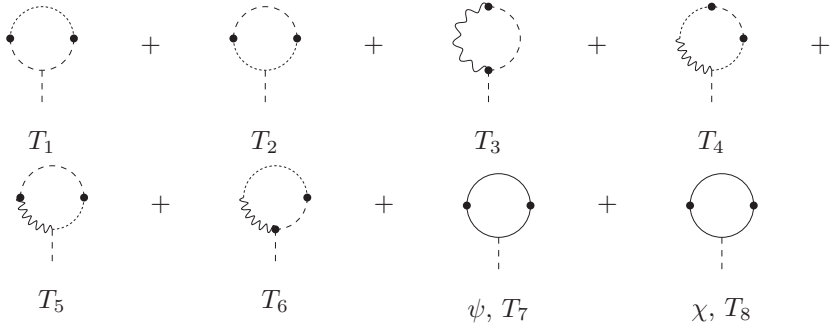}
     }
  }
\caption[]{Divergent one-loop contributions to the $\rho$ tadpole with dependence on $b_\mu$.}
\label{Tadpole-Rho}
\end{figure}

The last Green function to be considered in this group is the $\rho$ tadpole. Its superficial degree of divergence is given by $\mathcal{D}=3-\sum c_i V_i^{SR}$ and, thus, it is possible to have divergent contributions up to third order in $b_\mu$. However, only the second order part survives after considering all the graphs. The second order in $b_\mu$ divergent graphs are depicted in Fig. \ref{Tadpole-Rho}. When summed, they will give us the result
\be
T^\Lambda=\frac{i}{16 \pi^2} v b^2  \left[-g^2(C-D)^2-g'^2(C'-D')^2+ \kappa^2 \lambda-3 \kappa^2 e'^2\right] \frac{1}{\epsilon}.
\ee
The coefficients of the divergent pieces of the three Green functions match, as was stated in eq. (\ref{mass2}).

\subsection{The third group}

The renormalization of the next set of terms to be considered is a kind of consistency test. For the term
\be
e'^2A_\mu A^\mu \phi^* \phi
\ee
of eq. (\ref{model2}), we have the counterterm
\be
\mathcal{L}_{CT3}=\delta_3 e'^2A_\mu A^\mu \phi^* \phi=\delta_3 \frac{e'^2}{2}A_\mu A^\mu \left(\rho^2 + \varphi^2 + 2v \rho + v^2\right).
\ee
So, we have four Green functions which should have their divergent parts interconnected. The interesting fact is that the four-point functions $AA\rho \rho$ and $AA \varphi \varphi$ have superficial degree of divergence given by
$\mathcal{D}=-\sum c_i V_i^{SR}$. This means that any graph with a super-renormalizable vertex is finite and that the divergent part of these Green functions does not depend on the vector $b_\mu$. At the same time, for the three-point function $AA\rho$, a divergent graph with one super-renormalizable vertex is possible. However, it is not possible to construct such a divergent graph with one insertion of $b_\mu$ and no other super-renormalizable vertex.

Thus, we are left with the vacuum polarization tensor $AA$, which, for consistency, should not have a divergent part dependent on $b_\mu$. The contributions which are linear in $b_\mu$ have already been considered in section III in the calculation of the induced finite CFJ term. The diagrams which represent the superficially divergent contributions to this two-point function with two insertions of $b_\mu$ are shown in Fig. \ref{A-A} and their result are presented in appendix B. The total divergent part is null, as expected.

\begin{figure}[h!]
\unitlength1cm
\centerline{\hbox{
     \epsffile{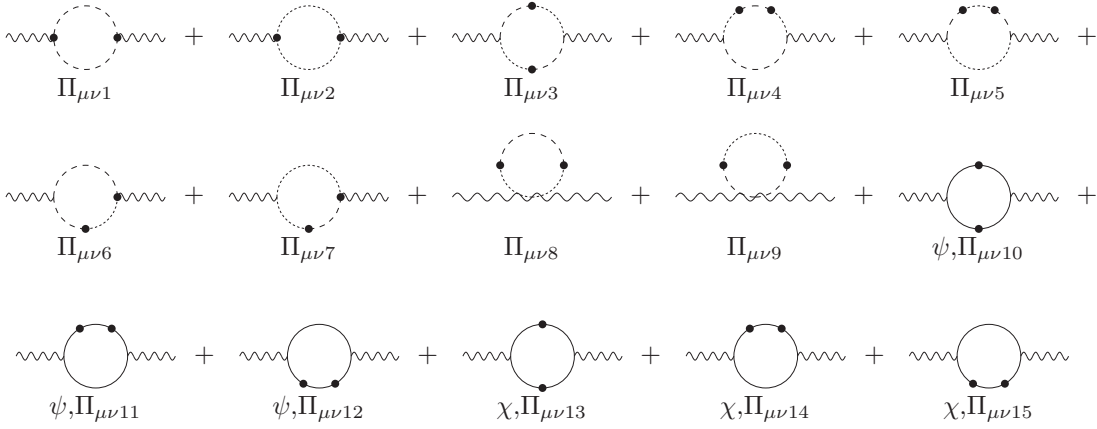}
     }
  }
\caption[]{Divergent one-loop contributions to the photon self-energy with dependence on $b_\mu$.}
\label{A-A}
\end{figure}

\subsection{Corrections to the fermionic lines}

\begin{figure}[h!]
\unitlength1cm
\centerline{\hbox{
     \epsffile{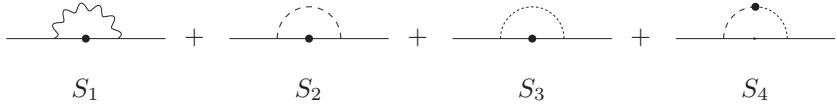}
     }
  }
\caption[]{Divergent one-loop contributions to the fermion self-energy with dependence on $b_\mu$.}
\label{Psi-Psi}
\end{figure}

The fermionic self-energies have the superficial degree of divergence given by $\mathcal{D}=1-\sum c_i V_i^{SR}$. Thus, it is possible to have divergent contributions of first order in $b_\mu$. The one-loop diagrams that represent these corrections are shown in Fig. \ref{Psi-Psi}, which are the same for the fermions $\psi$ and $\chi$. The final results are given by
\be
S^\Lambda_\psi=-\frac{i}{16 \pi^2}\bs \left\{\left[e^2M^2C+ \frac{g^2}{2}(D-1)\right]P_L+\left[e^2N^2D+\frac{g^2}{2}(C+1)\right]P_R\right\}\frac{1}{\epsilon}
\ee
and
\be
S^\Lambda_\chi=-\frac{i}{16 \pi^2}\bs \left\{\left[e^2N^2C'+ \frac{g'^2}{2}(D'-1)\right]P_L+\left[e^2M^2D'+\frac{g'^2}{2}(C'+1)\right]P_R\right\}\frac{1}{\epsilon}.
\ee
The results above have the general form of the coupling of the fermions to the background field already present in the Lagrangian density of the model.

\subsection{The mixed $\rho \varphi$ line}

\begin{figure}[h!]
\unitlength1cm
\centerline{\hbox{
     \epsffile{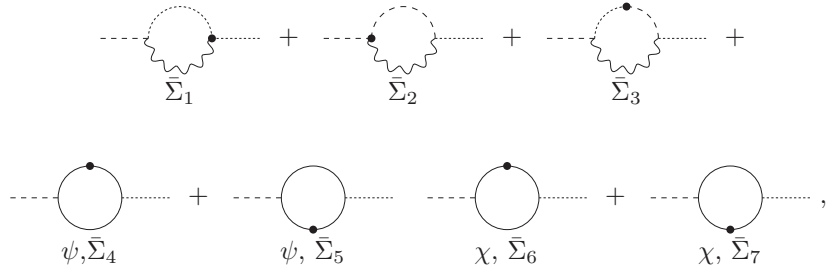}
     }
  }
\caption[]{Divergent one-loop contributions to the two-point function $\rho \varphi$ with dependence on $b_\mu$.}
\label{Rho-Phi}
\end{figure}

Finally, we have a divergent Lorentz-breaking part in the mixed two-point function $\rho \varphi$, which in principle could be quadratic in $b_\mu$. However, only the linear piece is non-zero, since the contributions cancel out in the case of fermionic loops and, in the absence of fermions, it is not possible to construct a second order graph in $b_\mu$ without another super-renormalizable vertex. The divergent Lorentz-violating graphs yielding this contribution are displayed in Fig. \ref{Rho-Phi}. The total result is given by
\be
\bar \Sigma^\Lambda=\frac{1}{8 \pi^2}(b \cdot p)\left\{g^2(C-D)+g'^2(C'-D')-2  \kappa e'^2\right\} \frac{1}{\epsilon}.
\ee
The divergent contribution above is perfectly absorbed by the the terms already present in the Lagrangian density of the model.

\subsection{The necessity of the Lorentz-breaking part in the covariant derivative}

To end this section, we comment on the necessity of including the Lorentz breaking part in the covariant derivative of the complex scalar field. Let us consider that the parameter $\kappa$ is zero. In this situation, we do not have the terms
\be
e' \kappa b_\mu A^\mu\left( \rho^2 + \varphi^2 + 2 v \rho + v^2\right)+ \kappa b^\mu \left(\rho \partial_\mu \varphi - \varphi \partial_\mu \rho\right)
\label{kappa-terms}
\ee
in the Lagrangian density of the model. Nevertheless, in the quantum computations corresponding to the first group of terms and to the mixed two-point function $\rho\varphi$, we obtain divergent contributions from other sectors, even if $\kappa=0$. Thus, the terms of (\ref{kappa-terms}) should be included by hand if they were not generated from the covariant derivative.

\section{Appendix B - Divergent $b_\mu$-dependent one-loop corrections}

In the integrals below, we write a fictitious mass $m_\varphi^2$ for the Goldstone boson which will not interfere in the calculation of the divergent parts. If the finite part is calculated, the limit $m_\varphi^2 \to 0$ should be taken.

The $b_\mu$-dependent divergent contributions to the one-point function in $A_\mu$ are given by:
\bq
T_{\mu 1}+T_{\mu 2}+T_{\mu 3}&=&e' \kappa \left\{-b_\mu\int\frac{d^4k}{(2 \pi)^4}\frac{1}{\left(k^2-m_\rho^2\right)}
-b_\mu\int\frac{d^4k}{(2 \pi)^4}\frac{1}{\left(k^2-m_\varphi^2\right)} \right.\nonumber \\
&+& \left. 4b^\rho \int\frac{d^4k}{(2 \pi)^4}\frac{k_\rho k_\mu}{\left(k^2-m_\rho^2\right)\left(k^2-m_\varphi^2\right)}\right\}=0,
\eq
\be
T_{\mu 4}=-4v^2 \kappa e'^3 b_\mu \int\frac{d^4k}{(2 \pi)^4}\frac{1}{\left(k^2-m_\rho^2\right)\left(k^2-m_A^2\right)}=-\frac{i}{4 \pi^2}v^2 \kappa e'^3 b_\mu \frac{1}{\epsilon}+ \mbox{FT},
\ee
\be
T_{\mu 5}=-4v^2 \kappa e'^3 b^\rho \int\frac{d^4k}{(2 \pi)^4}\frac{k^2 k_\rho k_\mu}{\left(k^2-m_\rho^2\right)\left(k^2-m_A^2\right)(k^2-m_\varphi^2)^2}
=-\frac{i}{16 \pi^2}v^2 \kappa e'^3 b_\mu \frac{1}{\epsilon} +\mbox{FT},
\ee
\be
T_{\mu 6}=-v^2 \kappa e'^3 b_\mu \int\frac{d^4k}{(2 \pi)^4}\frac{k^2}{\left(k^2-m_A^2\right)(k^2-m_\varphi^2)^2}=-\frac{i}{16 \pi^2}v^2 \kappa e'^3 b_\mu \frac{1}{\epsilon} +\mbox{FT},
\ee
\be
T_{\mu 7}=4v^2 \kappa e'^3 b^\rho \int\frac{d^4k}{(2 \pi)^4}\frac{k_\rho k_\mu}{\left(k^2-m_\rho^2\right)\left(k^2-m_A^2\right)(k^2-m_\varphi^2)}=\frac{i}{16 \pi^2}v^2 \kappa e'^3 b_\mu \frac{1}{\epsilon} +\mbox{FT},
\ee
\be
T_{\mu 8}=4v^2 \kappa e'^3 b^\rho \int\frac{d^4k}{(2 \pi)^4}\frac{k_\rho k_\mu}{\left(k^2-m_\rho^2\right)\left(k^2-m_A^2\right)(k^2-m_\varphi^2)}=\frac{i}{16 \pi^2}v^2 \kappa e'^3 b_\mu \frac{1}{\epsilon} +\mbox{FT},
\ee
\be
T_{\mu 9}= -e \int\frac{d^4k}{(2 \pi)^4}\frac{N_{\mu 1}}{(k^2-m_\psi^2)^2}= -\frac{i}{8 \pi^2}v^2 e g^2 (M-N)(D-C) b_\mu \frac{1}{\epsilon} +\mbox{FT}
\ee
and
\be
T_{\mu 10}= -e \int\frac{d^4k}{(2 \pi)^4}\frac{N_{\mu 2}}{(k^2-m_\chi^2)^2}-\frac{i}{8 \pi^2}v^2 e g'^2 (N-M)(D'-C') b_\mu \frac{1}{\epsilon} +\mbox{FT},
\ee
with
\be
N_{\mu 1}= \mbox{tr}{\left\{\gamma_\mu (M P_L+N P_R)(\ks+m_\psi)\bs(C P_L+D P_R)(\ks+m_\psi)\right\}}
\ee
and
\be
N_{\mu 2}= \mbox{tr}{\left\{\gamma_\mu (N P_L+M P_R)(\ks+m_\chi)\bs(C' P_L+D' P_R)(\ks+m_\chi)\right\}}
\ee
and where FT stands for finite terms.

The contributions to the divergent part of the mixed two-point function $A \rho$ and which depend on $b_\mu$ are given by
\be
F_{\mu1}=- \frac 32 \kappa e'v\lambda b_\mu \int\frac{d^4k}{(2 \pi)^4}\frac{1}{(k^2-m_\rho^2)^2}+ \mbox{FT}=- \frac {3i}{32 \pi^2} \kappa e'v\lambda b_\mu  \frac{1}{\epsilon}+\mbox{FT},
\ee
\be
F_{\mu2}=- \frac 12 \kappa e'v\lambda b_\mu \int\frac{d^4k}{(2 \pi)^4}\frac{1}{(k^2-m_\varphi^2)^2}+ \mbox{FT}=- \frac {i}{32 \pi^2} \kappa e'v\lambda b_\mu  \frac{1}{\epsilon}+\mbox{FT},
\ee
\be
F_{\mu3}=6 e'v\lambda \kappa b^\rho \int\frac{d^4k}{(2 \pi)^4}\frac{k_\mu k_\rho}{(k^2-m_\rho^2)^2(k^2-m_\varphi^2)}+ \mbox{FT}
=\frac {3i}{32 \pi^2} \kappa e'v\lambda b_\mu  \frac{1}{\epsilon}+\mbox{FT},
\ee
\be
F_{\mu4}=2 \kappa e'v\lambda b^\rho \int\frac{d^4k}{(2 \pi)^4}\frac{k_\mu k_\rho}{(k^2-m_\rho^2)(k^2-m_\varphi^2)^2}+ \mbox{FT}
= \frac {i}{32 \pi^2} \kappa e'v\lambda b_\mu  \frac{1}{\epsilon}+\mbox{FT},
\ee
\be
F_{\mu5}=-4 \kappa e'^3 v b_\mu \int\frac{d^4k}{(2 \pi)^4}\frac{1}{(k^2-m_\rho^2)(k^2-m_A^2)}+ \mbox{FT}
=- \frac {i}{4 \pi^2} \kappa e'^3 v b_\mu  \frac{1}{\epsilon}+\mbox{FT},
\ee
\be
F_{\mu6}=4 \kappa e'^3 v b^\rho \int\frac{d^4k}{(2 \pi)^4}\frac{k_\mu k_\rho}{(k^2-m_\rho^2)(k^2-m_\varphi^2)(k^2-m_A^2)}+ \mbox{FT}
=\frac {i}{16 \pi^2} \kappa e'^3 v b_\mu  \frac{1}{\epsilon}+\mbox{FT},
\ee
\be
F_{\mu7}=4 \kappa e'^3 v b^\rho \int\frac{d^4k}{(2 \pi)^4}\frac{k_\mu k_\rho}{(k^2-m_\rho^2)(k^2-m_\varphi^2)(k^2-m_A^2)}+ \mbox{FT}
=\frac {i}{16 \pi^2} \kappa e'^3 v b_\mu  \frac{1}{\epsilon}+\mbox{FT},
\ee
\be
F_{\mu 8}=-4 \kappa e'^3 v b_\mu \int\frac{d^4k}{(2 \pi)^4}\frac{1}{(k^2-m_\rho^2)(k^2-m_A^2)}+ \mbox{FT}
=-\frac {i}{4 \pi^2} \kappa e'^3 v b_\mu  \frac{1}{\epsilon}+\mbox{FT},
\ee
\be
F_{\mu9}=-4 \kappa e'^3 v b^\rho \int\frac{d^4k}{(2 \pi)^4}\frac{k^2k_\mu k_\rho}{(k^2-m_\rho^2)(k^2-m_\varphi^2)^2(k^2-m_A^2)}+ \mbox{FT}
=-\frac {i}{16 \pi^2} \kappa e'^3 v b_\mu  \frac{1}{\epsilon}+\mbox{FT},
\ee
\be
F_{\mu10}=-4 \kappa e'^3 v b^\rho \int\frac{d^4k}{(2 \pi)^4}\frac{k^2k_\mu k_\rho}{(k^2-m_\rho^2)(k^2-m_\varphi^2)^2(k^2-m_A^2)}+ \mbox{FT}
=-\frac {i}{16 \pi^2} \kappa e'^3 v b_\mu  \frac{1}{\epsilon}+\mbox{FT},
\ee
\be
F_{\mu11}=2 \kappa e'^3 v b_\mu \int\frac{d^4k}{(2 \pi)^4}\frac{k^2}{(k^2-m_\varphi^2)^2(k^2-m_A^2)}+ \mbox{FT}
=\frac {i}{8 \pi^2} \kappa e'^3 v b_\mu  \frac{1}{\epsilon}+\mbox{FT},
\ee
\be
F_{\mu12}=4 \kappa e'^3 v b^\rho \int\frac{d^4k}{(2 \pi)^4}\frac{k_\mu k_\rho}{(k^2-m_\rho^2)(k^2-m_\varphi^2)(k^2-m_A^2)}+ \mbox{FT}
=\frac {i}{16 \pi^2} \kappa e'^3 v b_\mu  \frac{1}{\epsilon}+\mbox{FT},
\ee
\be
F_{\mu13}=4 \kappa e'^3 v b^\rho \int\frac{d^4k}{(2 \pi)^4}\frac{k_\mu k_\rho}{(k^2-m_\rho^2)(k^2-m_\varphi^2)(k^2-m_A^2)}+ \mbox{FT}
=\frac {i}{16 \pi^2} \kappa e'^3 v b_\mu  \frac{1}{\epsilon}+\mbox{FT},
\ee
\be
F_{\mu 14}=-\frac{ge}{\sqrt{2}} \int\frac{d^4k}{(2 \pi)^4} \frac{R_{\mu 1}}{(k^2-m_\psi^2)^2[(k+p)^2-m_\psi^2]}
=-\frac {i}{16 \pi^2} g^2 e v (M-N)(D-C)b_\mu  \frac{1}{\epsilon}+\mbox{FT},
\ee
\be
F_{\mu 15}=-\frac{ge}{\sqrt{2}} \int\frac{d^4k}{(2 \pi)^4} \frac{R_{\mu 2}}{(k^2-m_\psi^2)[(k+p)^2-m_\psi^2]^2}
=-\frac {i}{16 \pi^2} g^2 e v (M-N)(D-C)b_\mu  \frac{1}{\epsilon}+\mbox{FT},
\ee
\be
F_{\mu 16}=-\frac{g'e}{\sqrt{2}} \int\frac{d^4k}{(2 \pi)^4} \frac{R_{\mu 3}}{(k^2-m_\chi^2)^2[(k+p)^2-m_\chi^2]}
=-\frac {i}{16 \pi^2} g'^2 e v (N-M)(D'-C')b_\mu  \frac{1}{\epsilon}+\mbox{FT}
\ee
and
\be
F_{\mu 17}=-\frac{g'e}{\sqrt{2}} \int\frac{d^4k}{(2 \pi)^4}\frac{R_{\mu 4}}{(k^2-m_\chi^2)[(k+p)^2-m_\chi^2]^2}
=-\frac {i}{16 \pi^2} g'^2 e v (N-M)(D'-C')b_\mu  \frac{1}{\epsilon}+\mbox{FT},
\ee
with
\be
R_{\mu 1}=\mbox{tr}\left\{(\ks+\ps+m_\psi)\gamma_\mu(M P_L+ N P_R)(\ks+m_\psi)\bs(C P_L+ D P_R)(\ks+m_\psi)\right\},
\ee
\be
R_{\mu 2}=\mbox{tr}\left\{(\ks+\ps+m_\psi)\bs(C P_L+ D P_R)(\ks+\ps+m_\psi)\gamma_\mu(M P_L+ N P_R)(\ks+m_\psi)\right\},
\ee
\be
R_{\mu 3}=\mbox{tr}\left\{(\ks+\ps+m_\chi)\gamma_\mu(N P_L+ M P_R)(\ks+m_\chi)\bs(C' P_L+ D' P_R)(\ks+m_\chi)\right\}
\ee
and
\be
R_{\mu 4}= \mbox{tr}\left\{(\ks+\ps+m_\chi)\bs(C' P_L+ D' P_R)(\ks+\ps+m_\chi)\gamma_\mu(N P_L+ M P_R)(\ks+m_\chi)\right\}.
\ee

For the divergent graphs of Fig. \ref{Rho-Rho-A}, we have the results:
\be
\Gamma_{\mu 1}=- \frac 32 \kappa e'\lambda b_\mu \int\frac{d^4k}{(2 \pi)^4}\frac{1}{(k^2-m_\rho^2)^2}+\mbox{FT}
=-\frac{3i}{32 \pi^2} \kappa e' \lambda b_\mu \frac{1}{\epsilon}+\mbox{FT},
\ee
\be
\Gamma_{\mu 2}=- \frac 12 \kappa e'\lambda b_\mu \int\frac{d^4k}{(2 \pi)^4}\frac{1}{(k^2-m_\varphi^2)^2}+\mbox{FT}
=-\frac{i}{32 \pi^2} \kappa e' \lambda b_\mu \frac{1}{\epsilon}+\mbox{FT},
\ee
\be
\Gamma_{\mu 3}= 6 \kappa e'\lambda b^\rho \int\frac{d^4k}{(2 \pi)^4}\frac{k_\mu k_\rho}{(k^2-m_\rho^2)^2(k^2-m_\varphi^2)}+ \mbox{FT}
=\frac{3i}{32 \pi^2} \kappa e' \lambda b_\mu \frac{1}{\epsilon}+\mbox{FT},
\ee
\be
\Gamma_{\mu 4}=2 \kappa e'\lambda b^\rho \int\frac{d^4k}{(2 \pi)^4}\frac{k_\mu k_\rho}{(k^2-m_\rho^2)(k^2-m_\varphi^2)^2}+ \mbox{FT}
=\frac{i}{32 \pi^2} \kappa e' \lambda b_\mu \frac{1}{\epsilon}+\mbox{FT},
\ee
\be
\Gamma_{\mu 5}=2 \kappa e'^3b_\mu \int\frac{d^4k}{(2 \pi)^4}\frac{k^2}{(k^2-m_\varphi^2)^2(k^2-m_A^2)}+\mbox{FT}
=\frac{i}{8 \pi^2} \kappa e'^3 b_\mu \frac{1}{\epsilon}+\mbox{FT},
\ee
\be
\Gamma_{\mu 6}=-8 \kappa e'^3 b^\rho \int\frac{d^4k}{(2 \pi)^4}\frac{k^2k_\mu k_\rho}{(k^2-m_\rho^2)(k^2-m_\varphi^2)^2(k^2-m_A^2)}+\mbox{FT}
=-\frac{i}{8 \pi^2} \kappa e'^3 b_\mu \frac{1}{\epsilon}+\mbox{FT},
\ee
\be
\Gamma_{\mu 7}=8 \kappa e'^3 b^\rho \int\frac{d^4k}{(2 \pi)^4}\frac{k_\mu k_\rho}{(k^2-m_\rho^2)(k^2-m_\varphi^2)(k^2-m_A^2)}+ \mbox{FT}
=\frac{i}{8 \pi^2} \kappa e'^3 b_\mu \frac{1}{\epsilon}+\mbox{FT},
\ee
\be
\Gamma_{\mu 8}=-8 \kappa e'^3 b_\mu \int\frac{d^4k}{(2 \pi)^4}\frac{1}{(k^2-m_\rho^2)(k^2-m_A^2)}+\mbox{FT}
=-\frac{i}{2 \pi^2} \kappa e'^3 b_\mu \frac{1}{\epsilon}+\mbox{FT},
\ee
\be
\Gamma_{\mu 9}=8 \kappa e'^3 b^\rho \int\frac{d^4k}{(2 \pi)^4}\frac{k_\mu k_\rho}{(k^2-m_\rho^2)(k^2-m_\varphi^2)(k^2-m_A^2)}+\mbox{FT}
=\frac{i}{8 \pi^2} \kappa e'^3 b_\mu \frac{1}{\epsilon}+\mbox{FT},
\ee
\be
\Gamma_{\mu10}+\Gamma_{\mu 11}+\Gamma_{\mu 12}= eg^2 \int\frac{d^4k}{(2 \pi)^4}\frac{G_{\mu 1}}{(k^2-m_\psi^2)^4}+ \mbox{FT}
=-\frac{i}{8 \pi^2} g^2 e (M-N)(D-C) b_\mu \frac{1}{\epsilon}+\mbox{FT}
\ee
and
\be
\Gamma_{\mu13}+\Gamma_{\mu 14}+\Gamma_{\mu 15}= eg'^2\int\frac{d^4k}{(2 \pi)^4}\frac{G_{\mu 2}}{(k^2-m_\chi^2)^4}+ \mbox{FT}
=-\frac{i}{8 \pi^2} g'^2 e (N-M)(D'-C') b_\mu \frac{1}{\epsilon}+\mbox{FT},
\ee
with
\bq
&&G_{\mu 1}=\mbox{tr}\left\{\gamma_\mu(M P_L+ N P_R) \ks \bs(C P_L+D P_R)\ks \ks \ks \right.\nonumber \\
&&\left.+\gamma_\mu(M P_L+N P_R) \ks \ks \bs (C P_L +D P_R) \ks \ks +\right. \nonumber \\
&&\left.+ \gamma_\mu (M P_L +N P_R) \ks \ks \ks \bs (C P_L+ D P_R) \ks \right\}
\eq
and
\bq
&&G_{\mu 2}=\mbox{tr}\left\{\gamma_\mu(N P_L+ M P_R) \ks \bs(C' P_L+D' P_R)\ks \ks \ks \right. \nonumber \\
&&\left.+\gamma_\mu(N P_L+M P_R) \ks \ks \bs (C' P_L +D' P_R) \ks \ks +\right. \nonumber \\
&&\left.+ \gamma_\mu (N P_L +M P_R) \ks \ks \ks \bs (C' P_L+ D' P_R) \ks \right\}.
\eq

For the graphs of Fig. \ref{Rho-Rho}, we have the following results:
\be
\Sigma_1=-4 \kappa^2 e'^2 b^2\int\frac{d^4k}{(2 \pi)^4}\frac{1}{(k^2-m_\rho^2)(k^2-m_A^2)}+\mbox{FT}=-\frac{i}{4 \pi^2} \kappa^2 e'^2b^2 \frac{1}{\epsilon}+ \mbox{FT},
\ee
\be
\Sigma_2=8 \kappa^2 e'^2 b^\alpha b^\beta\int\frac{d^4k}{(2 \pi)^4}\frac{k_\alpha k_\beta}{(k^2-m_\rho^2)(k^2-m_\varphi^2)(k^2-m_A^2)}
+ \mbox{FT}=\frac{i}{8 \pi^2} \kappa^2 e'^2b^2 \frac{1}{\epsilon}+ \mbox{FT},
\ee
\be
\Sigma_3=-4 \kappa^2e'^2 b^\alpha b^\beta\int\frac{d^4k}{(2 \pi)^4}\frac{k^2 k_\alpha k_\beta}{(k^2-m_\rho^2)(k^2-m_\varphi^2)^2(k^2-m_A^2)}
+ \mbox{FT}=-\frac{i}{16 \pi^2} \kappa^2 e'^2 b^2 \frac{1}{\epsilon}+ \mbox{FT},
\ee
\be
\Sigma_4=3 \lambda \kappa^2 b^\alpha b^\beta\int\frac{d^4k}{(2 \pi)^4}\frac{k_\alpha k_\beta}{(k^2-m_\rho^2)^2(k^2-m_\varphi^2)}
+ \mbox{FT}=\frac{3i}{64 \pi^2} \lambda \kappa^2 b^2 \frac{1}{\epsilon}+ \mbox{FT},
\ee
\be
\Sigma_5= \lambda \kappa^2 b^\alpha b^\beta\int\frac{d^4k}{(2 \pi)^4}\frac{k_\alpha k_\beta}{(k^2-m_\rho^2)^2(k^2-m_\varphi^2)}
+ \mbox{FT}=\frac{i}{64 \pi^2} \lambda \kappa^2 b^2 \frac{1}{\epsilon}+ \mbox{FT},
\ee
\bq
\Sigma_6+\Sigma_7+\Sigma_8&=&\frac{g^2}{2}\int\frac{d^4k}{(2 \pi)^4}\frac{1}{(k^2-m_\psi^2)^4}
\mbox{tr}\left\{ \ks \bs (CP_L+DP_R)\ks \bs(CP_L+DP_R)\ks \ks +\right. \nonumber \\
&+& \left. \ks \bs (CP_L+DP_R)\ks \ks \bs(CP_L+DP_R)\ks + \ks \ks \bs (CP_L+DP_R)\ks \bs(CP_L+DP_R)\ks \right\} \nonumber \\
&+& \mbox{FT}=-\frac{i}{16 \pi^2} g^2 b^2 \frac{1}{\epsilon}+ \mbox{FT}
\eq
and
\bq
\Sigma_9+\Sigma_{10}+\Sigma_{11}&=&\frac{g'^2}{2}\int\frac{d^4k}{(2 \pi)^4}\frac{1}{(k^2-m_\chi^2)^4}
\mbox{tr}\left\{ \ks \bs (CP_L+DP_R)\ks \bs(CP_L+DP_R)\ks \ks +\right. \nonumber \\
&+& \left. \ks \bs (CP_L+DP_R)\ks \ks \bs(CP_L+DP_R)\ks + \ks \ks \bs (CP_L+DP_R)\ks \bs(CP_L+DP_R)\ks \right\} \nonumber \\
&+& \mbox{FT}=-\frac{i}{16 \pi^2} g'^2 b^2 \frac{1}{\epsilon}+ \mbox{FT}.
\eq

The results for the amplitudes represented by the diagrams of Fig. \ref{Tadpole-Rho} read
\be
T_1= 3 v \lambda \kappa^2 b^\alpha b^\beta\int\frac{d^4k}{(2 \pi)^4}\frac{k_\alpha k_\beta}{(k^2-m_\rho^2)^2(k^2-m_\varphi^2)} = \frac {3i}{64 \pi^2} v \lambda \kappa^2 b^2 \frac {1}{\epsilon}+ \mbox{FT},
\ee
\be
T_2= v \lambda \kappa^2 b^\alpha b^\beta\int\frac{d^4k}{(2 \pi)^4}\frac{k_\alpha k_\beta}{(k^2-m_\rho^2)(k^2-m_\varphi^2)^2} = \frac {i}{64 \pi^2} v \lambda \kappa^2 b^2 \frac {1}{\epsilon}+ \mbox{FT},
\ee
\be
T_3= -4 v e'^2 \kappa^2 b^2\int\frac{d^4k}{(2 \pi)^4}\frac{1}{(k^2-m_\rho^2)(k^2-m_\varphi^2)} = -\frac {i}{4 \pi^2} v e'^2 \kappa^2 b^2 \frac {1}{\epsilon}+ \mbox{FT},
\ee
\be
T_4= -4 v e'^2 \kappa^2 b^\alpha b^\beta\int\frac{d^4k}{(2 \pi)^4}\frac{k^2 k_\alpha k_\beta}{(k^2-m_\rho^2)(k^2-m_\varphi^2)^2(k^2-m_A^2)} = -\frac {i}{16 \pi^2} v e'^2 \kappa^2 b^2 \frac {1}{\epsilon}+ \mbox{FT},
\ee
\be
T_5= 4 v e'^2 \kappa^2 b^\alpha b^\beta\int\frac{d^4k}{(2 \pi)^4}\frac{k_\alpha k_\beta}{(k^2-m_\rho^2)(k^2-m_\varphi^2)(k^2-m_A^2)} = \frac {i}{16 \pi^2} v e'^2 \kappa^2 b^2 \frac {1}{\epsilon}+ \mbox{FT},
\ee
\be
T_6= 4 v e'^2 \kappa^2 b^\alpha b^\beta\int\frac{d^4k}{(2 \pi)^4}\frac{k_\alpha k_\beta}{(k^2-m_\rho^2)(k^2-m_\varphi^2)(k^2-m_A^2)} = \frac {i}{16 \pi^2} v e'^2 \kappa^2 b^2 \frac {1}{\epsilon}+ \mbox{FT},
\ee
\bq
&&T_7= \frac{g}{\sqrt{2}}\int\frac{d^4k}{(2 \pi)^4}\frac{1}{(k^2-m_\psi^2)^3}\mbox{tr}\left\{(\ks+m_\psi)\bs(CP_L+DP_R)(\ks+m_\psi)\bs(CP_L+DP_R)(\ks+m_\psi)\right\} \nonumber \\
&&= -\frac {i}{16 \pi^2} v g^2 (C-D)^2 b^2 \frac {1}{\epsilon}+ \mbox{FT}
\eq
and
\bq
&&T_8= \frac{g'}{\sqrt{2}}\int\frac{d^4k}{(2 \pi)^4}\frac{1}{(k^2-m_\chi^2)^3}\mbox{tr}\left\{(\ks+m_\chi)\bs(C'P_L+D'P_R)(\ks+m_\chi)\bs(C'P_L+D'P_R)(\ks+m_\chi)\right\} \nonumber \\
&&= -\frac {i}{16 \pi^2} v g'^2 (C'-D')^2 b^2 \frac {1}{\epsilon}+ \mbox{FT}.
\eq

The diagrams displayed in Fig. \ref{A-A} have the following results:
\be
\Pi_{\mu \nu 1}= 2 e'^2 \kappa^2 b_\mu b_\nu\int\frac{d^4k}{(2 \pi)^4}\frac{1}{(k^2-m_\rho^2)^2}
+ \mbox{FT}=\frac{i}{8 \pi^2} e'^2 \kappa^2 b_\mu b_\nu \frac{1}{\epsilon}+ \mbox{FT},
\ee
\be
\Pi_{\mu \nu 2}= 2 e'^2 \kappa^2 b_\mu b_\nu\int\frac{d^4k}{(2 \pi)^4}\frac{1}{(k^2-m_\varphi^2)^2}
+ \mbox{FT}=\frac{i}{8 \pi^2} e'^2 \kappa^2 b_\mu b_\nu \frac{1}{\epsilon}+ \mbox{FT},
\ee
\be
\Pi_{\mu \nu 3}=16 e'^2 \kappa^2 b^\alpha b^\beta \int\frac{d^4k}{(2 \pi)^4}\frac{k_\alpha k_\beta k_\mu k_\nu}{(k^2-m_\rho^2)^2(k^2-m_\varphi^2)^2}
+\mbox{FT}=\frac{i}{24 \pi^2} \kappa^2 \left(b^2 g_{\mu \nu}+ 2 b_\mu b_\nu\right) \frac{1}{\epsilon}+ \mbox{FT},
\ee
\be
\Pi_{\mu \nu 4}=16 e'^2 \kappa^2 b^\alpha b^\beta \int\frac{d^4k}{(2 \pi)^4}\frac{k_\alpha k_\beta k_\mu k_\nu}{(k^2-m_\rho^2)^2(k^2-m_\varphi^2)^2}
+ \mbox{FT}=\frac{i}{24 \pi^2} \kappa^2 \left(b^2 g_{\mu \nu}+ 2 b_\mu b_\nu\right) \frac{1}{\epsilon}+ \mbox{FT},
\ee
\be
\Pi_{\mu \nu 5}=16 e'^2 \kappa^2 b^\alpha b^\beta \int\frac{d^4k}{(2 \pi)^4}\frac{k_\alpha k_\beta k_\mu k_\nu}{(k^2-m_\rho^2)^2(k^2-m_\varphi^2)^2}
+ \mbox{FT}=\frac{i}{24 \pi^2} \kappa^2 \left(b^2 g_{\mu \nu}+ 2 b_\mu b_\nu\right) \frac{1}{\epsilon}+ \mbox{FT},
\ee
\be
\Pi_{\mu \nu 6}=-16 e'^2 \kappa^2 b_\mu b^\alpha  \int\frac{d^4k}{(2 \pi)^4}\frac{k_\alpha k_\nu}{(k^2-m_\rho^2)^2(k^2-m_\varphi^2)}
+ \mbox{FT}=-\frac{i}{4 \pi^2} \kappa^2 b_\mu b_\nu \frac{1}{\epsilon}+ \mbox{FT},
\ee
\be
\Pi_{\mu \nu 7}=-16 e'^2 \kappa^2 b_\mu b^\alpha  \int\frac{d^4k}{(2 \pi)^4}\frac{k_\alpha k_\nu}{(k^2-m_\rho^2)(k^2-m_\varphi^2)^2}
+ \mbox{FT}=-\frac{i}{4 \pi^2} \kappa^2 b_\mu b_\nu \frac{1}{\epsilon}+ \mbox{FT},
\ee
\be
\Pi_{\mu \nu 8}=-4 e'^2 g_{\mu \nu} \kappa^2 b^\alpha b^\beta  \int\frac{d^4k}{(2 \pi)^4}\frac{k_\alpha k_\beta}{(k^2-m_\rho^2)(k^2-m_\varphi^2)^2}
+ \mbox{FT}=-\frac{i}{16 \pi^2} \kappa^2 b^2 g_{\mu \nu} \frac{1}{\epsilon}+ \mbox{FT},
\ee
\be
\Pi_{\mu \nu 9}=-4 e'^2 g_{\mu \nu} \kappa^2 b^\alpha b^\beta  \int\frac{d^4k}{(2 \pi)^4}\frac{k_\alpha k_\beta}{(k^2-m_\rho^2)^2(k^2-m_\varphi^2)}
+ \mbox{FT}=-\frac{i}{16 \pi^2} \kappa^2 b^2 g_{\mu \nu} \frac{1}{\epsilon}+ \mbox{FT},
\ee
\bq
&&\Pi_{\mu \nu 10}+\Pi_{\mu \nu 11}+\Pi_{\mu \nu 12}= -e^2 \int\frac{d^4k}{(2 \pi)^4}\frac{1}{(k^2-m_\psi^2)^4} \times\nonumber \\
&& \times \mbox{tr}\left\{ \gamma_\mu(MP_L+NP_R)\ks \bs (CP_L+DP_R)\ks \bs(CP_L+DP_R)\ks\gamma_\nu(MP_L+NP_R) \ks +\right. \nonumber \\
&&+ \left. \gamma_\mu(MP_L+NP_R)\ks \bs (CP_L+DP_R)\ks \gamma_\nu(MP_L+NP_R) \ks \bs(CP_L+DP_R)\ks \right. +\nonumber \\
&&+ \left. \gamma_\mu(MP_L+NP_R)\ks \gamma_\nu(MP_L+NP_R) \ks \bs (CP_L+DP_R)\ks \bs(CP_L+DP_R)\ks \right\}
+ \mbox{FT} =0
\eq
and
\bq
&&\Pi_{\mu \nu 13}+\Pi_{\mu \nu 14}+\Pi_{\mu \nu 15}= -e^2 \int\frac{d^4k}{(2 \pi)^4}\frac{1}{(k^2-m_\chi^2)^4} \times \nonumber \\
&&\times \mbox{tr}\left\{ \gamma_\mu(NP_L+MP_R)\ks \bs (C'P_L+D'P_R)\ks \bs(C'P_L+D'P_R)\ks\gamma_\nu(NP_L+MP_R) \ks +\right. \nonumber \\
&&+ \left. \gamma_\mu(NP_L+MP_R)\ks \bs (C'P_L+D'P_R)\ks \gamma_\nu(NP_L+MP_R) \ks \bs(C'P_L+D'P_R)\ks \right. +\nonumber \\
&&+ \left. \gamma_\mu(NP_L+MP_R)\ks \gamma_\nu(NP_L+MP_R) \ks \bs (C'P_L+D'P_R)\ks \bs(C'P_L+D'P_R)\ks \right\}
+ \mbox{FT}=0.
\eq

The corrections corresponding to the graphs of Fig. \ref{Psi-Psi} for the fermion $\psi$ are given by
\bq
S_{\psi1}&=&-e^2\int\frac{d^4k}{(2 \pi)^4}\frac{\gamma^\rho \ks \bs \ks \gamma_\rho \left(M^2C P_L+N^2D P_R\right)}{(k^2-m_\psi^2)^2(k^2-m_A^2)}+ \mbox{FT} \nonumber \\
&=&-\frac{i}{16 \pi^2} e^2 \bs \left(M^2C P_L+N^2D P_R\right)\frac{1}{\epsilon}+ \mbox{FT},
\eq
\bq
S_{\psi2}&=&\frac{g^2}{2}\int\frac{d^4k}{(2 \pi)^4}\frac{\ks \bs \ks \left(D P_L + C P_R\right)}{(k^2-m_\psi^2)^2(k^2-m_\rho^2)}+ \mbox{FT}\nonumber \\
&=&-\frac{i}{64 \pi^2} g^2 \bs \left(D P_L + C P_R\right)\frac{1}{\epsilon}+ \mbox{FT},
\eq
\bq
S_{\psi3}&=&\frac{g^2}{2}\int\frac{d^4k}{(2 \pi)^4}\frac{\ks \bs \ks \left(D P_L + C P_R\right)}{(k^2-m_\psi^2)^2(k^2-m_\varphi^2)}+ \mbox{FT}\nonumber \\
&=&-\frac{i}{64 \pi^2} g^2 \bs \left(D P_L + C P_R\right)\frac{1}{\epsilon}+ \mbox{FT}
\eq
and
\bq
S_{\psi4}&=&-2g^2 \gamma^\alpha \gamma_5 b^\beta \int\frac{d^4k}{(2 \pi)^4}\frac{k_\alpha k_\beta}{(k^2-m_\psi^2)(k^2-m_\rho^2)(k^2-m_\varphi^2)}+ \mbox{FT}\nonumber \\
&=&-\frac{i}{32 \pi^2} g^2 \bs \left(-P_L + P_R\right)\frac{1}{\epsilon}+ \mbox{FT}.
\eq

Finally, the divergent Lorentz-violating contributions to the mixed two-point function depicted in Fig. \ref{Rho-Phi} have the results
\be
\bar \Sigma_1=2ie'^2 \kappa b^\alpha \int\frac{d^4k}{(2 \pi)^4}\frac{(2p+k)_\alpha}{(k^2-m_A^2)[(k+p)^2-m_\varphi^2]}=-\frac{3}{16 \pi^2}e'^2 \kappa (b \cdot p) \frac{1}{\epsilon} + \mbox{FT},
\ee
\be
\bar \Sigma_2=2ie'^2 \kappa b^\alpha \int\frac{d^4k}{(2 \pi)^4}\frac{(2p+k)_\alpha}{(k^2-m_A^2)[(k+p)^2-m_\rho^2]}=-\frac{3}{16 \pi^2}e'^2 \kappa (b \cdot p) \frac{1}{\epsilon} + \mbox{FT},
\ee
\be
\bar \Sigma_3=-2ie'^2 \kappa b^\alpha \int\frac{d^4k}{(2 \pi)^4}\frac{(2p+k)^2(p+k)_\alpha}{(k^2-m_A^2)[(k+p)^2-m_\varphi^2][(k+p)^2-m_\rho^2]}=\frac{1}{8 \pi^2}e'^2 \kappa (b \cdot p) \frac{1}{\epsilon} + \mbox{FT},
\ee
\bq
\bar \Sigma_4&=&-i\frac{g^2}{2} b^\alpha \int\frac{d^4k}{(2 \pi)^4}\frac{\mbox{tr}\left\{(\ks + \ps) \gamma_5 \ks \bs (C P_L + D P_R)\ks\right\}}{(k^2-m_\psi^2)^2[(k+p)^2-m_\psi^2]} + \mbox{FT} \nonumber \\
&=&\frac{1}{16 \pi^2}g^2 (C-D) (b \cdot p) \frac{1}{\epsilon} + \mbox{FT},
\eq
\bq
\bar \Sigma_5&=&-i\frac{g^2}{2} b^\alpha \int\frac{d^4k}{(2 \pi)^4}\frac{\mbox{tr}\left\{(\ks + \ps) \bs (C P_L + D P_R)(\ks + \ps) \gamma_5 \ks\right\}}{(k^2-m_\psi^2)[(k+p)^2-m_\psi^2]^2} + \mbox{FT} \nonumber \\
&=&\frac{1}{16 \pi^2}g^2 (C-D) (b \cdot p) \frac{1}{\epsilon} + \mbox{FT},
\eq
\bq
\bar \Sigma_6&=&-i\frac{g'^2}{2} b^\alpha \int\frac{d^4k}{(2 \pi)^4}\frac{\mbox{tr}\left\{(\ks + \ps) \gamma_5 \ks \bs (C' P_L + D' P_R)\ks\right\}}{(k^2-m_\chi^2)^2[(k+p)^2-m_\chi^2]} + \mbox{FT} \nonumber \\
&=&\frac{1}{16 \pi^2}g'^2 (C'-D') (b \cdot p) \frac{1}{\epsilon} + \mbox{FT},
\eq
and
\bq
\bar \Sigma_7&=&-i\frac{g'^2}{2} b^\alpha \int\frac{d^4k}{(2 \pi)^4}\frac{\mbox{tr}\left\{(\ks + \ps) \bs (C' P_L + D' P_R)(\ks + \ps) \gamma_5 \ks\right\}}{(k^2-m_\chi^2)[(k+p)^2-m_\chi^2]^2} + \mbox{FT} \nonumber \\
&=&\frac{1}{16 \pi^2}g'^2 (C'-D') (b \cdot p) \frac{1}{\epsilon} + \mbox{FT},
\eq


\begin{thebibliography}{1}

\bibitem{Gross-Jackiw} David J. Gross, R. Jackiw, Phys. Rev. D6, 477 (1972).

\bibitem{data-exp} V. A. Kostelecky and Neil Russell, Rev. Mod. Phys. 83, 11 (2011).

\bibitem{jackiw} S.M. Carroll, G. B. Field and R. Jackiw, Phys. Rev. D41, 1231 (1990).

\bibitem{kostelecky1}  D. Colladay and V. A. Kostelecky, Phys. Rev. D55, 6760 (1997).

\bibitem{kostelecky2}  D. Colladay and V. A. Kostelecky, Phys. Rev. D58, 116002 (1998).

\bibitem{coleman1}  S. Coleman and S. L. Glashow, Phys. Lett. B405, 249 (1997).

\bibitem{coleman2} S. Coleman and S. L. Glashow, Phys. Rev. D59, 116008 (1999).

\bibitem{CS1} R. Jackiw and V. A. Kostelecky, Phys. Rev. Lett. 82, 3572 (1999).

\bibitem{CS2} J. M. Chung and B. K. Chung, Phys. Rev. D63, 105015 (2001).

\bibitem{CS3} J.M. Chung, Phys.Rev. D60, 127901 (1999).

\bibitem{CS4} W. F. Chen, Phys. Rev. D60, 085007 (1999).

\bibitem{CS5} M. Perez-Victoria, Phys. Rev. Lett. 83, 2518 (1999).

\bibitem{CS6} M. Perez-Victoria, J. High. Energy Phys. 0104, 032 (2001).

\bibitem{CS7} A. P. Ba\^eta Scarpelli, M. Sampaio, M. C. Nemes, and B. Hiller, Phys. Rev. D64, 046013 (2001).

\bibitem{CS8} T. Mariz, J.R. Nascimento, E. Passos, R.F. Ribeiro and F.A. Brito, J. High. Energy Phys. 0510 019 (2005).

\bibitem{CS9} B. Altschul, Phys. Rev. D70, 101701 (2004).

\bibitem{CS10} A.P. Ba\^eta Scarpelli, M. Sampaio, M.C. Nemes, B. Hiller, Eur. Phys. J. C56, 571 (2008).

\bibitem{Gomes} M. Gomes, J. R. Nascimento, A. Yu. Petrov and A. J. da Silva, Phys. Rev. D81, 045018 (2010).

\bibitem{Cleber} L. C. T. Brito, H. G. Fargnoli, A.P. Ba\^eta Scarpelli, Phys. Rev. D87, 12, 125023 (2013).

\bibitem{amb-free} F. A. Brito, J. R. Nascimento, E. Passos, A. Yu. Petrov, Phys. Lett. B664, 112 (2008).

\bibitem{Adriano} A. Cherchiglia, M. Sampaio, and M. Nemes, Int. J. Mod. Phys. A26, 2591 (2011), arXiv:1008.1377 [hep-th].

\bibitem{ouranom} A. P. Ba\^eta Scarpelli, T. Mariz, J. R. Nascimento, A. Yu. Petrov, arXiv: 1505.04047 [hep-th].

\end{thebibliography}
\end{document}